\documentstyle[11pt,psfig,amssymb,amsmath]{article}

\begin{document}
\title{Plateau Instability of Liquid Crystalline Cylinder\\
in Magnetic Field}

\author{Leonid G. Fel and Yoram Zimmels\\
\\Department of Civil and Environmental Engineering, Technion, \\Haifa 
32000, Israel}
\maketitle

\def\be{\begin{equation}}
\def\ee{\end{equation}}
\def\p{\prime}

\begin{abstract}
The capillary instability of a LC cylinder in magnetic field is considered 
using an energy approach. The boundary problem is solved in the linear 
approximation of the anisotropy $\chi_a$ of the magnetic susceptibility $\chi$.
The effect of anisotropy, in the region $1\gg|\chi|>|\chi_a|\gg \chi^2$, 
can 
be strong enough to counteract and even reverse the tendency of the field to 
enhance stabilization by enlarging the cut--off wave number $k_s$ beyond 
the conventional one set by Rayleigh.
\end{abstract}

\vskip .5cm
\noindent 
\begin{tabbing}
\hspace{.35in} {\sf Key words}:\hspace{.1in}  \=  Plateau Instability, Nematic 
Liquid Crystal, 
Magnetic Field, \\
               \>Anisotropy of Susceptibility.
\end{tabbing}
\vskip 2.5cm
\centerline{e-mail: lfel@techunix.technion.ac.il}
\newpage

\noindent
\section{Introduction}
\label{introd1}
Theoretical predictions of the continuum theory of the nematic liquid 
crystals (LC) were successfully confirmed in many experimental observations 
\cite{Genn74}. One of the most studied effect is the influence of an external 
field on the orientational distribution of the LC director field ${\bf n}({\bf 
r})$. The physics involved in a competition between the mechanical and field 
forces that can be developed in LC media, calls for careful analysis. Well 
known example is the Freedericksz effect \cite{Genn74}, observed when nematic 
LC cell with initial uniform 
distribution of ${\bf n}({\bf r})$ is subjected to an external magnetic field. 
In many cases, the stabilizing elastic forces compete with the destabilizing 
magnetic field giving rise to a critical phenomenon. However, the critical 
phenomena in LC can be sustained even if both the elastic and magnetic fields 
are defined as stabilizing. Such critical phenomena exist due to the effect of 
surface tension which tends to minimize the surface area by distorting the 
initial shape of the system.

The nematicity of LC's, being a source for elastic properties, results in 
enhancement of stability of LC jets \cite{felzim02}, as compared to ordinary 
liquids. A similar stability enhancement appears in ordinary liquid jets with 
isotropic magnetic permeability (see \cite{Ros85} and \cite{felzim03}) when they
are subjected to an external magnetic field. Unlike the elasticity, the external
field has a critical value beyond which instability of the jet is completely 
suppressed for all disturbance wavelengths \cite{felzim03}. Nematic LC's are 
usually {\it anisotropic diamagnetics} with positive anisotropy $\chi_a$ of the 
magnetic susceptibility \cite{Genn74}. This poses an additional challenge with 
respect to the above mentioned phenomena. It is reflected by the extra terms 
in the LC hydrodynamics due to orientational interaction between the magnetic 
field ${\bf H}$ and LC director ${\bf n}$.

The static version of capillary instability in liquid jets is known as the 
Plateau instability in the liquid cylinders \cite{felzim02}, \cite{felzim03}. It
dates back to the classical works of J. Plateau \cite{Plat73} who defined the
problem of finding a surface of liquid with a minimal area $S$ given its 
boundary $\partial \Omega$ at fixed volume. The problem relates to the  
principle of minimum free energy at equilibrium. Further generalization is 
called for if the excess free energy $W$, of the cylinder, comprises different 
types of energy that reflect a more complex structure of the liquid (e.g. 
elasticity \cite{felzim02} {\it etc}), as well as its capacity to interact with 
external fields.

The purpose of this work is to extend the theory of Plateau instability in LC 
cylinders \cite{felzim02} so as to include the effect of static magnetic fields.
Here the motivation is both experimental and theoretical. Experimentally, the 
question is how to set the initial orientation of director {\bf n} collinear 
with the LC cylinder axis. A weak magnetic field can serve to this end. 
Theoretically, the framework outlined in \cite{felzim02} can be extended so as 
to incorporate the influence of external fields on the evolution and stability 
of the LC cylinder. In particular, the Plateau instability is studied with 
respect to the effect of the magnetic anisotropy of the LC cylinder.

\section{Free energy of LC cylinder in the presence of magnetic field}
\label{planmagnet1}
Consider {\it an isothermal incompressible LC cylinder in a uniform magnetic 
field} ${\bf H}_0$ that is applied in free space along the cylinder axis. We 
assume a rigid boundary condition (BC) where the director is tangentially anchored at the free 
surface of the LC cylinder. The magnetic susceptibility tensor $\widehat\chi$ 
of the LC is assumed anisotropic, symmetric, and independent of the magnetic 
field. In the reference frame related to the cylinder axis its diagonal terms 
are $\chi_{\parallel},\chi_{\perp}$ while its off--diagonal term is $\chi_an_
zn_r$, where $n_z,n_r$ are the axial and radial components of director {\bf n}, 
respectively.
\begin{figure}[h]
\vspace{-1cm}
\centerline{\psfig{figure=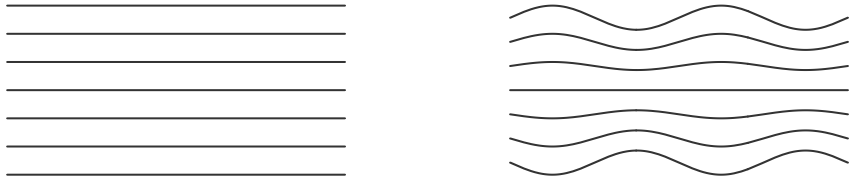,height=6cm,width=12cm}}
\vspace{-1.5cm}
\caption{Undisturbed ({\it left}) and disturbed ({\it right}) homotropic LC
cylinder subjected to an axial and uniform external magnetic field.}
\label{plannn1}
\end{figure}

\noindent
When the LC cylinder (assumed long compared to its diameter) is undisturbed 
the total free energy $F^0$ of the system is defined by
\begin{equation}
F^0={\cal E}_s^0-\chi_{\parallel}\frac{\mu_0H_0^2}{2}\cdot
\int_{\Omega_{cyl}^0}dv\;.\label{stat1}
\end{equation}
where the integral represents the volume $\pi R^2L$ of the undisturbed 
cylinder which is enclosed by the surface $\partial \Omega_{cyl}^0$. The
term ${\cal E}_s^0=\sigma \int_{\partial \Omega_{cyl}^0}ds=2\pi \sigma R L$
stands for the surface free energy of the undisturbed cylinder, where $R,L$
and $\sigma$ denote its radius, length and surface tension respectively, and  
$\mu_0$ is permeability of a free space. We specify the commonly used harmonic 
deformation of the cylinder as
$r(z)=R+\zeta_0\cos kz$, where $k=2\pi/\Lambda$, $\Lambda$ being the 
disturbance wavelength. Let the extent of deformation be characterized by 
a length $\zeta_0$, such that $\zeta_0/R=\epsilon\ll 1$.

Deformation of the cylinder shape changes the magnetic field ${\bf H}({\bf r})$ 
over all space ${\mathbb R}^3$, while the director field ${\bf n}({\bf r})$ 
is changed only within the internal domain $\Omega_{cyl}$. Following 
Plateau, we assume conservation of the cylinder volume
\begin{equation}
\int_{\Omega_{cyl}^0}dv=\int_{\Omega_{cyl}}dv\;.\label{stat2a}
\end{equation}  
The total free energy $F$ of the disturbed cylinder takes the following form,
\begin{equation}
F={\cal E}_s+{\cal E}_n+{\cal E}_H^{\sf in}+{\cal E}_H^{\sf ex}\;.
\label{distrib0}
\end{equation}
The first term in (\ref{distrib0}), which stands for the interfacial energy 
of the disturbed cylinder, is classically known due to Plateau \cite{Plat73}
\begin{eqnarray}
{\cal E}_s=\sigma_0 \int_{\partial \Omega_{cyl}} ds=
2\pi \sigma RL+\sigma\frac{\pi \zeta_0^2}{2 R}L\left(k^2R^2-1\right)\;.
\label{distrib11}
\end{eqnarray}
The second term in (\ref{distrib0}) is due to the elastic deformation of the 
director field ${\bf n}({\bf r})$, and in the single elastic approximation 
is given by
\begin{eqnarray}
{\cal E}_n=\frac{K}{2}\int_{\Omega_{cyl}}\left({\rm div}^2{\bf n}+
{\rm rot}^2{\bf n}\right)dv\;,\label{distrib12}
\end{eqnarray}
where $K$ is the elastic modulus. 
The last two terms in (\ref{distrib0}) correspond to the effect of the 
magnetic fields in the internal $\Omega_{cyl}$ and external ${\mathbb R}^3
\setminus\Omega_{cyl}$ domains
\begin{eqnarray}
{\cal E}_H^{\sf in}=\frac{\mu_0}{2}\int_{\Omega_{cyl}}H_0^2dv
-\frac{\mu_0}{2}\int_{\Omega_{cyl}}\mu_{jk}\;{}^{\sf in}H_j\;{}^{\sf 
in}H_kdv,\;\;\;\;{\cal E}_H^{\sf ex}=\frac{\mu_0}{2}
\int_{{\mathbb R}^3\setminus\Omega_{cyl}}H_0^2dv-\frac{\mu_0}{2}
\int_{{\mathbb R}^3\setminus \Omega_{cyl}}\;^{\sf ex}{\bf H}^2({\bf r})dv,
\label{distrib00}
\end{eqnarray}
where ${}^{\sf in}{\bf H}({\bf r})$ and ${}^{\sf ex}{\bf H}({\bf r})$ are the 
internal and external magnetic fields, respectively. If the deviations of 
the director ${\bf n}={\bf n}^0+{\bf n}^1$ from its initial orientation 
${\bf n}^0$ along the $z$ direction are small, then 
\begin{eqnarray}
n^0_r=0,\;n^0_z=1\;,\;\;1\gg n^1_r\gg |n^1_z| \sim \left(n^1_x\right)^2\;.
\label{frank3d}
\end{eqnarray}
The magnetic energy density in the second term of ${\cal E}_H^{\sf in}$ (
scaled by $\mu_0/2$) reads
\begin{eqnarray}
\mu_{jk}\;{}^{\sf in}H_j\;{}^{\sf in}H_k&=&
\left(1+\chi_{\parallel}\right)\left(\;{}^{\sf in}H_z\right)^2+
\left(1+\chi_{\perp}\right)\left(\;{}^{\sf in}H_r\right)^2+
2\chi_a n_r^1n_z^0\;{}^{\sf in}H_r\;{}^{\sf in}H_z\;,\label{distrib1}
\end{eqnarray}
where $\mu_{jk}$ is the LC relative permeability tensor: $\mu_{zz}=1+\chi_
{\parallel},\;\mu_{rr}=1+\chi_{\perp},\;\mu_{zr}=\chi_an_r^1n_z^0$. The 
excess free energy $W$ of the system is defined as,
\begin{equation}
W=F-F^0\;.\label{stat5}   
\end{equation}
From the mathematical standpoint, the variational problem for minimization
of $W$, supplemented with constraint (\ref{stat2a}) for all smooth surfaces 
$\partial \Omega_{cyl}$, is known as the isoperimetric problem. The cylinder 
instability can be studied assuming small perturbation in its  shape. In this 
case  the Plateau problem becomes solvable in closed form. The fields $^{\sf 
in}{\bf H}({\bf r})$ and $^{\sf ex}{\bf H}({\bf r})$, which must satisfy 
Maxwell equations, can be presented as small perturbations of ${\bf H}_0$,
\begin{equation}
^{\sf in}{\bf H}({\bf r})={\bf H}_0+^{\sf in}\!\!{\bf H}^1({\bf r})=
\left(H_0+^{\sf in}\!\!H_z^1,\;^{\sf in}H_r^1\right)\;,\;\;\;  
^{\sf ex}{\bf H}({\bf r})={\bf H}_0+^{\sf ex}\!\!{\bf H}^1({\bf r})=
\left(H_0+^{\sf ex}\!\!H_z^1,\;^{\sf ex}H_r^1\right)\;,\label{stat7}
\end{equation}
where according to the assumption $\epsilon\ll 1$ the following
approximations apply 
\begin{equation}
\left\{^{\sf in}H_r^1,\;^{\sf ex}H_r^1,\;^{\sf in}H_z^1,
\;^{\sf ex}H_z^1\right\}=
\left\{^{\sf in}h_r^1,\;^{\sf ex}h_r^1,\;^{\sf in}h_z^1,
\;^{\sf ex}h_z^1\right\}\times \epsilon H_0\;.\label{stat8}
\end{equation}
The dimensionless fields $^{\sf in,ex}h_{r,z}^1(r,z)$ are dependent on the
coordinates as indicated By virtue of translational invariance of the problem
\begin{eqnarray}
^{\sf in,ex}h_{r,z}^1(r,z+\Lambda)=\;^{\sf in,ex}h_{r,z}^1(r,z)\label{stat9}
\end{eqnarray}
we set $L=\Lambda$ and evaluate the free energy per unit wave length. 
Substituting (\ref{stat1}), (\ref{distrib11})--(\ref{distrib1}) and 
(\ref{stat7}) into (\ref{stat5}) we obtain in the $\epsilon^2$--approximation 
\begin{eqnarray}
\frac{1}{L}W=\epsilon^2\frac{\pi}{2}\sigma R\left(k^2R^2-1\right)+
\frac{K}{2L}\int_{\Omega_{cyl}}\left({\rm div}^2{\bf n}+
{\rm rot}^2{\bf n}\right)dv-\frac{\mu_0}{2L}\;U\;,\label{stat8a}   
\end{eqnarray}
where the magnetic part was calculated in Appendix \ref{appendix1}
\begin{eqnarray}
U&=&\int_{\Omega_{cyl}}\left\{(1+\chi_{\parallel})\left(^{\sf 
in}H_z^1\right)^2+
(1+\chi_{\perp})\left(^{\sf in}H_r^1\right)^2\right\}dv+\int_{{\mathbb R}^3
\setminus \Omega_{cyl}}\left\{\;\left(^{\sf ex}H_z^1\right)^2+
\left(^{\sf ex}H_r^1\right)^2\right\}dv+\nonumber\\
&&2H_0\left(\chi_a \int_{\Omega_{cyl}}n_r n_z \;^{\sf in}H_r^1 dv+   
(1+\chi_{\parallel})\int_{\Omega_{cyl}}\;^{\sf in}H_z^1dv+\int_{{\mathbb R}^3
\setminus \Omega_{cyl}}\;^{\sf ex}H_z^1dv\right)\;.\label{stat9}
\end{eqnarray}
\section{Boundary problem and its solution}
\label{isotrop1}
The magnetostatics of the disturbed LC cylinder is governed by Maxwell 
equations for the internal $^{\sf in}{\bf H}({\bf r})$ and external 
$^{\sf ex}{\bf H}({\bf r})$ magnetic fields and the Euler--Lagrange equation 
apply for the director field ${\bf n}({\bf r})$,
\begin{eqnarray}
&&{\rm rot}\;^{\sf in}{\bf H}={\rm rot}\;^{\sf ex}{\bf H}=0\;,\;\;\;\;
{\rm div}\;^{\sf in}{\bf B}={\rm div}\;^{\sf ex}{\bf B}=0\;,\label{max1}\\
&&\left\{\frac{\partial }{\partial r}\;\frac{\partial }
{\partial (\partial_r n_r)}+\frac{\partial }{\partial z}\;\frac{\partial }   
{\partial (\partial_z n_r)}-\frac{\partial }{\partial n_r}\right\}
\left({\cal E}_n+{\cal E}_H^{\sf in}\right)=0\;,\;\;\;
\partial_x=\frac{\partial }{\partial x}\;.\label{max1a}
\end{eqnarray}
where $^{\sf in}{\bf B}$ and $^{\sf ex}{\bf B}$ denote internal and 
external magnetic inductions, respectively
\begin{eqnarray}
^{\sf in}B_j=\mu_0\mu_{jk}\;^{\sf in}H_k\;,\;\;\mu_{jk}=\left(1+\chi_{\perp}
\right)\delta_{jk}+\chi_a n_jn_k\;,\;\;\;^{\sf ex}B_j=\mu_0\;^{\sf ex}H_j\;,\;
\;\chi_a=\chi_{\parallel}-\chi_{\perp}\;.\label{induct1}
\end{eqnarray}
Equations (\ref{max1}), (\ref{max1a}) must be supplemented with boundary 
conditions (BC) at the interface $r=R$, 
\begin{eqnarray}
\langle {}^{\sf in}{\bf H},{\bf t}\rangle=
\langle {}^{\sf ex}{\bf H},{\bf t}\rangle\;,\;\;\;\;
\langle {}^{\sf in}{\bf B},{\bf e}\rangle=
\langle {}^{\sf ex}{\bf B},{\bf e}\rangle\;,\;\;\;\;
\langle {\bf e},{\bf n}\rangle=0\;,\label{max2}
\end{eqnarray}
where {\bf t} and {\bf e} stand for tangential and normal unit vectors to
the surface, respectively. Since the surface deformation is small,
linearization can be applied,
\begin{eqnarray}
t_r=-e_z=\partial \zeta/\partial z\;,\;\;\;t_z=e_r=
\sqrt{1-\left(\partial \zeta/\partial z\right)^2}\simeq 1\;.\label{max2a}
\end{eqnarray}
A standard way to solve the problem is to introduce a director potential 
$\Theta({\bf r})$ and two magnetic potentials $\Phi_{\sf in}({\bf r})$ and 
$\Phi_{\sf ex}({\bf r})$ as follows
\begin{eqnarray}
n^1_r=\frac{\partial \Theta}{\partial r}\;,\;\;\;
{}^{\sf in}{\bf H}^1({\bf r})=-\nabla\Phi_{\sf in}\;,\;\;\;
{}^{\sf ex}{\bf H}^1({\bf r})=-\nabla\Phi_{\sf ex}\;,\;\;\;
|\nabla\Phi_{\sf in}|, |\nabla\Phi_{\sf ex}|\ll H_0\;.\label{max2b}   
\end{eqnarray}
$\Phi_{\sf in}({\bf r})$ and $\Phi_{\sf ex}({\bf r})$ satisfy the first 
two equations in (\ref{max1}). The last two equations in (\ref{max1}) yield,
\begin{eqnarray}
\left(1+\chi_{\parallel}\right)\frac{\partial^2 \Phi_{\sf in}}{\partial z^2}+
\left(1+\chi_{\perp}\right)\Delta_{2} \Phi_{\sf in}=\chi_aH_0\frac{1}{r}\frac{
\partial }{\partial r}\left(r \frac{\partial \Theta}{\partial r}\right)\;,\;\;
\;\;\frac{\partial^2 \Phi_{\sf ex}}{\partial z^2}+
\Delta_2\Phi_{\sf ex}=0\;,\label{max3}
\end{eqnarray}
where $\Delta_2=\partial^2/\partial r^2+1/r \partial/\partial r$ is the 
two--dimensional Laplacian. The variational equation (\ref{max1a}) gives,
\begin{eqnarray}
K\left(\Delta_{2}-\frac{1}{r^2}+\frac{\partial^2 }{\partial z^2}\right)
\frac{\partial \Theta}{\partial r}-2\mu_0\chi_aH_0 \frac{\partial 
\Phi_{\sf in}}{\partial r}=0\;.\label{max3a}
\end{eqnarray}
Making use of the commutation rules 
$$
\left(\Delta_{2}-\frac{1}{r^2}+\frac{\partial^2 }{\partial z^2}\right)
\frac{\partial \Theta}{\partial r}=\frac{\partial }{\partial r}
\left(\Delta_{2}+\frac{\partial^2 }{\partial z^2}\right) \Theta\;,
$$
we finally arrive at
\begin{eqnarray}
&&\left(\Delta_{2}+\frac{\partial^2}{\partial z^2}\right)\Phi_{\sf ex}=0\;,\;\;
\left(\Delta_{2}+\alpha^2\frac{\partial^2 }{\partial z^2}\right)\Phi_{\sf in}=
(\alpha^2-1)H_0\Delta_{2}\Theta\;,\label{magn6}\\
&&\left(\Delta_{2}+\frac{\partial^2}{\partial z^2}\right)\Theta=
2\frac{\mu_0\chi_aH_0}{K}\Phi_{\sf in}\;,\;\;\;\;
0\leq \alpha^2-1=\frac{\chi_a}{1+\chi_{\perp}}\simeq \chi_a\ll 1\;.\nonumber
\end{eqnarray}
BC (\ref{max2}) can be reformulated as follows
\begin{eqnarray}
\frac{\partial \Phi_{\sf ex}}{\partial z}=\frac{\partial \Phi_{\sf 
in}}{\partial z}\;,\;\;\;\;\;
\frac{\partial \Phi_{\sf ex}}{\partial r}-\left(1+\chi_{\perp}\right)
\frac{\partial \Phi_{\sf in}}{\partial r}=\chi_{\perp}H_0\frac{\partial
\zeta}{\partial z}\;,\;\;\;\;\;\frac{\partial
\Theta}{\partial r}=\frac{\partial \zeta}{\partial z}\;.\label{bond3}
\end{eqnarray}
A weak decoupling of the equations (\ref{magn6}) makes it possible to solve 
the boundary problem in closed form. Assuming 
\begin{equation}
\left\{\Phi_{\sf in}(r,z),\Phi_{\sf ex}(r,z),\Theta(r,z)\right\}=\left\{
\phi_{\sf in}(r),\phi_{\sf ex}(r),\theta(r)\right\}\times\sin kz\label{contr0}
\end{equation}
we find
\begin{equation}
\left(\Delta_{2}-k^2\right)\phi_{\sf ex}=0\;,\;\;
\left(\Delta_{2}-\alpha^2k^2\right)\phi_{\sf in}=\chi_aH_0\Delta_{2}\theta\;,
\;\;\left(\Delta_{2}-k^2\right)\theta=2\frac{\mu_0\chi_aH_0}{K}\phi_{\sf 
in}\;,\label{contr1}
\end{equation}
The following BC exist at $r=R$
\begin{equation}
\phi_{\sf ex}=\phi_{\sf in}\;,\;\;\;
\left(1+\chi_{\perp}\right)\frac{\partial \phi_{\sf in}}{\partial r}-
\frac{\partial \phi_{\sf ex}}{\partial r}=\chi_{\perp}H_0k\zeta_0\;,\;\;\;\;\;
\frac{\partial \theta}{\partial r}=-k\zeta_0\;.\label{contr1a}
\end{equation}
The two last equations in (\ref{contr1}) can be represented through the 
determinant equation
\begin{eqnarray}
&&\left[\Delta_{2}^2-D_1\Delta_{2}+D_0\right]
\left(\begin{array}{c}\phi_{\sf in}\\\theta\end{array}\right)=
\left(\begin{array}{c}0\\0\end{array}\right)\;,\;\;\;
\mbox{where}\label{contr3}\\
&&D_1=(2+\chi_a)k^2+\chi_ag^2\;.\;\;D_0=(1+\chi_a)k^4\;,\;\;
g^2=2\frac{\mu_0\chi_aH_0^2}{K}\;.\label{contr4}
\end{eqnarray}
Factorization of the differential operator in (\ref{contr3}) gives
\begin{eqnarray}
&&\Delta_{2}^2-D_1\Delta_{2}+D_0=
\left(\Delta_{2}-l_1^2\right)\left(\Delta_{2}-l_2^2\right)\;,\;\;\;
\mbox{where}\;\;\;l_{1,2}^2=\frac{1}{2}\left(D_1\pm\sqrt{{\cal D}}\right),
\label{contr5}\\
&&{\cal D}=D_1^2-4D_0=\chi_a\left[\chi_a(k^4+g^4)+2(2+\chi_a)k^2g^2\right]\;.
\label{contr5a}
\end{eqnarray}
The  fundamental solutions of (\ref{contr3}) which are finite at $r=0$ are 
the following
\begin{equation}
\phi_{\sf in}(r)=c_1I_0(l_1r)+c_2I_0(l_2r),\;\;\;\theta(r)=b_1I_0(l_1r)+
b_2I_0(l_2r),\;\;\;\frac{\chi_aH_0}{1+\chi_{\perp}}b_j=\frac{l_j^2-
\alpha^2k^2}{l_j^2}c_j,\;j=1,2,\label{contr6}
\end{equation}
where $I_m(x)$ is a modified Bessel function of the 1st kind and order $m$, and 
$c_j,b_j$ are indeterminates. The first equation in (\ref{contr1}) produces a 
solution in the exterior domain which is finite at $r=\infty$
\begin{equation}
\phi_{\sf ex}(r)=c_3 K_0(kr)\;,\label{contr2}
\end{equation}
where $K_m(x)$ is a modified Bessel function of the 2nd kind and order $m$.
All indeterminates $c_1,c_2,c_3$ can be found from BC (\ref{contr1a}) by 
substitution therein the expressions (\ref{contr6}), (\ref{contr2}). 
Further simplification comes after substitution of $c_3=c_1I_0(l_1R)/K_0(kR)+
c_2I_0(l_2R)/K_0(kR)$ and using $1+\chi_{\perp}\simeq 1$
\begin{eqnarray}
\begin{array}{l}
c_1\left[l_1I_1(l_1R)+k\frac{K_1(kR)}{K_0(kR)}I_0(l_1R)\right]\\
c_1\left(\alpha^2k^2-l_1^2\right)\frac{I_1(l_1R)}{l_1}\end{array}
\begin{array}{c}+\\+\end{array}
\begin{array}{l}
c_2\left[l_2I_1(l_2R)+k\frac{K_1(kR)}{K_0(kR)}I_0(l_2R)\right]\\
c_2\left(\alpha^2k^2-l_2^2\right)\frac{I_1(l_2R)}{l_2}\end{array}
\begin{array}{c}=\\=\end{array}
\begin{array}{c}
\chi_{\perp}k\zeta_0H_0\\
\chi_ak\zeta_0H_0\end{array}\;,\label{contr71}
\end{eqnarray}
where the identities $I_0^{\prime}(x)=I_1(x)$ and $K_0^{\prime}(x)=-K_1(x)$ 
for the derivatives were used. Straightforward calculations give
\begin{eqnarray}
c_j=k\zeta_0H_0\;\frac{\Gamma_j}{\Gamma_0}\;,\;\;
b_j=k\zeta_0\;\frac{\Gamma_j}{\Gamma_0}
\frac{l_j^2-\alpha^2k^2}{\chi_al_j^2}\;,\;\;j=1,2\;,\label{contr8} 
\end{eqnarray}
where $\Gamma_j, j=0,1,2$ are determinants of $(2\times 2)$ matrices
\begin{eqnarray}
\Gamma_0&=&\alpha\left(l_1^2-l_2^2\right)I_1(l_1R)I_1(l_2R)+
\frac{kK_1(kR)}{K_0(kR)}\left[(\alpha l_1-l_2)I_0(l_1R)I_1(l_2R)-
(\alpha l_2-l_1)I_0(l_2R)I_1(l_1R)\right]\nonumber\\
-\Gamma_1&=&\left(\chi_{\parallel}l_2^2-\chi_{\perp}\alpha^2k^2\right)
\frac{I_1(l_2R)}{l_2}+\chi_ak\frac{K_1(kR)}{K_0(kR)}I_0(l_2R)\;,
\nonumber\\
\Gamma_2&=&\left(\chi_{\parallel}l_1^2-\chi_{\perp}\alpha^2k^2\right)
\frac{I_1(l_1R)}{l_1}+\chi_ak\frac{K_1(kR)}{K_0(kR)}I_0(l_1R)\;.\label{contr9}
\end{eqnarray}
\subsection{$\chi_a$--expansion of the solutions}
\label{expen}
The complexity of expressions (\ref{contr6}) in conjunction with (\ref{contr8})
and (\ref{contr9})  makes further evaluation of the problem excessively 
difficult. 
Therefore, we develop in this Section another approach for solution of the 
amplitude equations (\ref{contr1}) endowed with BC (\ref{contr1a}). Bearing in 
mind that for most nematic LCs the anisotropy is small $|\chi_a|<|\chi_{\perp}|,
|\chi_{\parallel}|$
\footnote{The characteristic magnitudes of the magnetic susceptibility $\chi$ 
and its anisotropy $\chi_a$ for the classical nematic LC's {\em 
4--metoxybenziliden--4--butilanilin (MBBA)} and {\em para--azoxyanisole (PAA)} 
can be found in \cite{DEJeu80} : $\chi_{\perp}\approx \chi_{\parallel}
\approx -10^{-5}$ and $\chi_a\approx 10^{-6}$.}
we seek the linear in $\chi_a$ representation of the functions $\phi_{\sf in}
(r)$ and $\theta(r)$ 
\begin{equation}
\phi_{\sf in}(r)=
{\overline \phi_{\sf in}}(r)+\chi_a {\widetilde \phi_{\sf in}}(r)\;,\;\;\;
\theta(r)={\overline \theta}(r)+\chi_a {\widetilde \theta}(r)\;.\label{exp1}
\end{equation}
The isotropic parts ${\overline \phi_{\sf in}}(r)$, ${\overline \theta}(r)$ 
together with the external potential $\phi_{\sf ex}$, satisfy the following 
equations
\begin{equation}
\left(\Delta_{2}-k^2\right)\phi_{\sf ex}=
\left(\Delta_{2}-k^2\right){\overline \phi_{\sf in}}=
\left(\Delta_{2}-k^2\right){\overline \theta}=0\;,
\label{exp2}
\end{equation}
supllemented with the BC at $r=R$
\begin{equation}
\phi_{\sf ex}={\overline \phi_{\sf in}}\;,\;\;\;
\left(1+\chi_{\perp}\right)\frac{\partial {\overline \phi_{\sf in}}}
{\partial r}-\frac{\partial \phi_{\sf ex}}{\partial r}=
\chi_{\perp}H_0k\zeta_0\;,\;\;\;\;\;
\frac{\partial {\overline \theta}}{\partial r}=-k\zeta_0\;.
\label{exp3}
\end{equation}
The solutions ${\overline \phi_{\sf in}}(r)$, ${\overline \theta}(r)$, 
$\phi_{\sf ex}(r)$ of (\ref{exp2}) were found in \cite{felzim02} and 
\cite{felzim03}
\begin{eqnarray}
{\overline \phi_{\sf in}}(r)=A_1I_0(kr)\;,\;\;\;
{\overline \theta}(r)=A_2I_0(kr)\;,\;\;\;
\phi_{\sf ex}(r)=A_3K_0(kr)\;,
\label{exp3a}
\end{eqnarray}
where the coefficients $C_i$ are given by,
\begin{eqnarray}
A_1=\zeta_0\chi_{\perp} H_0kR\frac{K_0(kR)}{T(kR,\chi_{\perp})}\;,\;\;\;
A_2=-\frac{\zeta_0}{I_1(kR)}\;,\;\;\;
A_3=\zeta_0\chi_{\perp} H_0kR\frac{I_0(kR)}{T(kR,\chi_{\perp})}\;,\label{exp3b}
\end{eqnarray}
and $T(x,a)=1+a x I_1(x)K_0(x)$. Henceforth, $T(kR,\chi_{\perp})\simeq 1$ in 
accordance with $\chi_{\perp}\ll 1$. The amplitude equations for the remaining 
functions ${\widetilde \phi_{\sf in}}(r)$ and ${\widetilde \theta}(r)$ can be 
found by inserting (\ref{exp1}) into (\ref{contr1}) and making use of 
(\ref{exp2}) 
\begin{eqnarray}
&&\left(\Delta_{2}-k^2\right){\widetilde \phi_{\sf in}}=A_{\phi}I_0(kr)\;,\;\;
\;\;\;\;\;A_{\phi}=(A_1+A_2H_0)k^2=-\frac{\zeta_0k^2H_0}{I_1(kR)}\;,\label{exp4}\\
&&\left(\Delta_{2}-k^2\right){\widetilde \theta}=A_{\theta}I_0(kr)\;,\;\;\;\;
\;A_{\theta}=2A_1\frac{\mu_0H_0}{K}=2\chi_{\perp}\zeta_0kRK_0(kR)
\frac{\mu_0H_0^2}{K}\;.\nonumber
\end{eqnarray}
The BC for ${\widetilde\phi_{\sf in}}(r)$ and ${\widetilde \theta}(r)$ is,
\begin{eqnarray}
{\widetilde \phi_{\sf in}}(R)={\widetilde \theta}(R)=0\;.\label{exp5}
\end{eqnarray}
After simple calculations (see Appendix \ref{appendix2}) we obtain
\begin{eqnarray}
{\widetilde \phi_{\sf in}}(kr)=\frac{A_{\phi}}{k^2}G(kr)I_0(kr)\;,\;\;\;
{\widetilde \theta}(kr)=\frac{A_{\theta}}{k^2}G(kr)I_0(kr)\;,\label{exp11}
\end{eqnarray}
where
\begin{eqnarray}
G(kr)=\frac{k^2}{4}\left(r^2-R^2\right)
+\frac{1}{2}\int^{kR}_{kr}\frac{I_1^2(y)}{I_0^2(y)}ydy\;,\;\;\;
\frac{G^{\prime}_r(kr)}{k}=\frac{kr}{2}\left(1-\frac{I_1^2(kr)}{I_0^2(kr)}
\right)\;.\label{exp12}
\end{eqnarray}
Recalling the definition (\ref{max2b}) of potentials $\Phi_{\sf in}({\bf r})$, 
$\Phi_{\sf ex}({\bf r})$ and $\Theta({\bf r})$ we get
\begin{eqnarray}
n^1_r(r,z)&=&kA_2I_1(kr)\left\{1+\chi_{a}\frac{A_{\theta}}{A_2k^2}
\left[G(kr)+\frac{G^{\prime}_r(kr)}{k}\frac{I_0(kr)}{I_1(kr)}
\right]\right\}\sin kz\;,\label{exp13a}\\
-{}^{\sf in}H_r^1(r,z)&=&kA_1I_1(kr)\left\{1+\chi_a\frac{A_{\phi}}{A_1k^2}\left
[G(kr)+\frac{G^{\prime}(kr)}{k}\frac{I_0(kr)}{I_1(kr)}\right]\right\}\sin kz\;,
\label{exp13b}\\
-{}^{\sf in}H_z^1(r,z)&=&kA_1I_0(kr)\left\{1+\chi_a\frac{A_{\phi}}{A_1k^2}
G(kr)\right\}\cos kz\;,\label{exp13c}\\ 
{}^{\sf ex}H_r^1(r,z)&=&kA_3K_1(kr)\sin kz\;,\;\;\;\;
{}^{\sf ex}H_z^1(r,z)=-kA_3K_0(kr)\cos kz\;.
\label{exp13}
\end{eqnarray}
where
\begin{eqnarray}
\chi_a\frac{A_{\theta}}{A_2k^2}\simeq -2\chi_a\chi_{\perp}\frac{\mu_0 H_0^2R^2}
{K}\frac{I_1(kR)K_0(kR)}{kR}\;,\;\;\;
\chi_a\frac{A_{\phi}}{A_1k^2}\simeq -\frac{\chi_a}{\chi_{\perp}}\frac{1}
{kRI_1(kR)K_0(kR)}\;.
\label{exp14}
\end{eqnarray}
Before going to further calculation of the excess free energy $W$ by 
(\ref{stat8a}), let us estimate the main contributions of the anisotropy 
$\chi_a$ to the distribution of the fields ${\bf n}(r,z)$ and ${}^{\sf 
in}{\bf H}^1(r,z)$ in accordance with (\ref{exp13}) and (\ref{exp14}). 

\begin{figure}[h]
\centerline{\psfig{figure=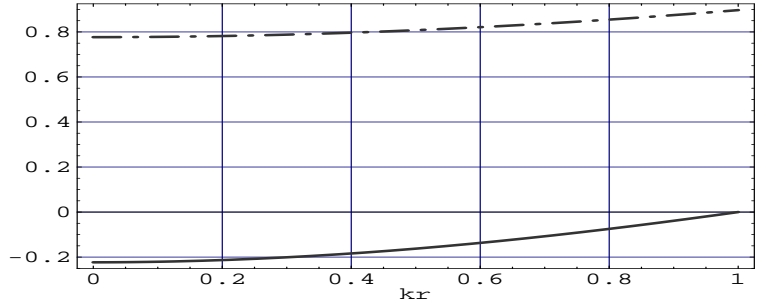,height=4cm,width=10cm}}
\caption{Plots of the functions $G(kr)$ ({\em solid curve}) and
$G(kr)+\frac{G^{\prime}(kr)}{k}\frac{I_0(kr)}{I_1(kr)}$ ({\em dashed curve}).}
\label{ggg0}
\end{figure}

First, as follows from Figure \ref{ggg0}, $G(kr)$ and $G(kr)+\frac{G^{\prime}
(kr)}{k}\frac{I_0(kr)}{I_1(kr)}$ which are both continuous monotone 

\noindent
growing funstions, are bounded as follows
\begin{eqnarray}
-0.21<G(kr)<0\;,\;\;\;\mbox{and}\;\;\;0.75<G(kr)+\frac{G^{\prime}(kr)}{k}
\frac{I_0(kr)}{I_1(kr)}<0.9\;.
\label{exp15}
\end{eqnarray}
In order to simplify further calculation we consider, henceforth, these 
functions as constant $N_1$ and $N_2$, respectively 
\begin{eqnarray}
G(kr)=N_1\;,\;\;\;G(kr)+\frac{G^{\prime}(kr)}{k}\frac{I_0(kr)}{I_1(kr)}=N_2\;,
\label{exp16}
\end{eqnarray}
The next simplification comes for $n^1_r(r,z)$. Indeed, bearing in mind 
(\ref{exp14}) we conclude that for the influence of $\chi_a$ on 
$n^1_r(r,z)$ to be significant a huge magnetic field $H_0$ is required 
\begin{eqnarray}
H_0>H_{\bullet}=\frac{1}{\sqrt{|\chi_a\chi_{\perp}|}}\times 
\frac{1}{R}\sqrt{\frac{K}{\mu_0}},
\label{exp16a}
\end{eqnarray}
The magnitude of this field can be as high as $10^8A/m$ for classical LC's 
with radius $R\approx 10 \mu m$. In fields which are significantly lower  
than $H_{\bullet}$, the behaviour of the director ${\bf n}({\bf r})$ is 
dictated primarily by competition between bulk elasticity and surface tension 
of the LC's, and goverened by dimensionless parameter 
$\varkappa=K/\sigma R$ \cite{felzim02}. 
Recasting (\ref{exp13a}), (\ref{exp13b}) and (\ref{exp13c}) gives
\begin{eqnarray}
n^1_r(r,z)&=&kA_2I_1(kr)\sin kz\;,\;\;{}^{\sf ex}H_r^1(r,z)=kA_3K_1(kr)\sin kz
\;,\;\;{}^{\sf ex}H_z^1(r,z)=-kA_3K_0(kr)\cos kz\;,\nonumber\\
-{}^{\sf in}H_r^1(r,z)&=&kA_1I_1(kr)\left(1+\chi_a\frac{A_{\phi}}
{A_1k^2}N_2\right)\sin kz\;,\;\;
-{}^{\sf in}H_z^1(r,z)=kA_1I_0(kr)\left(1+\chi_a\frac{A_{\phi}}{A_1k^2}N_1
\right)\cos kz.\nonumber
\end{eqnarray}
Now we are in position to calculate the magnetic part of the excess free energy 
$W$ according to (\ref{stat9}), in the limit $\chi_a<\chi_{\perp}\ll 1$ (see 
Appendix \ref{appendix1}). 
\begin{eqnarray}
-U=\pi L(\zeta_0H_0)^2\chi_{\perp}kR\frac{I_0(kR)}{I_1(kR)}\left(\chi_{\perp}
kRI_1(kR)K_0(kR)+2N_1\frac{\chi_a}{\chi_{\perp}}\right)\;.\label{integ21}
\end{eqnarray}
The elastic part of $W$ was found in \cite{felzim02}
\begin{eqnarray}
{\cal E}_n=\pi LKk^2\zeta_0^2\;.
\label{integ22}
\end{eqnarray}
Inserting (\ref{integ21}) and (\ref{integ22}) into (\ref{stat8a}) we get
\begin{eqnarray}
W=\frac{\pi L\sigma R}{2}\left(\frac{\zeta_0}{R}\right)^2 \cdot
f(kR,\chi,H_0)\;,\label{integ33}
\end{eqnarray}
where
\begin{eqnarray}
f(kR,\chi,H_0)=(kR)^2(1+2\varkappa)-1+\frac{\mu_0RH_0^2}{\sigma}\frac{kR 
I_0(kR)}{I_1(kR)}\left(\chi_{\perp}^2kR I_1(kR)K_0(kR)+2N_1\chi_a\right)\;.
\label{integ34}
\end{eqnarray}
All the terms in (\ref{integ34}), except the last one, describe the 
stabilization of a LC cylinder due to the existence of isotropic susceptibility 
irrespective of its sign \cite{felzim03} and due to the elasticity of the LC 
phase \cite{felzim02}. The influence of the last term in (\ref{integ34}) which 
accounts for the anisotropy of $\chi$ can be significant and even dominating. 
The latter occurs if 
\begin{eqnarray}
|\chi_a|\gg \chi_{\perp}^2\;,
\label{integ34a}
\end{eqnarray}
as indeed is the case in classical LC materials (MBBA, PAA). Here the physical 
situation changes completely. The cylinder is destabilized with the 
corresponding cut--off $k_sR$
\begin{eqnarray}
k_sR\simeq1+2\chi_a|N_1|\frac{\mu_0RH^2}{\sigma}\;.
\label{integ36}
\end{eqnarray}
The interesting property of (\ref{integ36}) is the fact that the cut--off $k_sR$ 
extends beyond the range $(0\leq k_sR\leq 1)$ of the classical Rayleigh 
instability. This kind of extension cannot be obtained as a field or elastic 
effects in the absence of anisotropy $\chi_a$ of the magnetic susceptibility.
\section{Conclusion}
\label{concl}
\begin{itemize}
\item 
The capillary instability of a LC cylinder in magnetic field is considered 
using an energy approach. The excess free energy, which includes terms due to 
surface, LC's elasticity, and magnetic field is used to find extremum 
conditions associated with instability. The boundary problem is solved and 
then expanded in terms of the anisotropy $\chi_a$ of the magnetic 
susceptibility.
\item
The excess magnetic free energy, which was founded to be a function of the 
isotropic susceptibility ($\chi$) squared proved to have an anisotropic part 
linear in $\chi_a$. This indicates that the effect of anisotropy can turn 
dominant provided that $1\gg|\chi_{\perp}|>|\chi_a|\gg\chi_{\perp}^2$. This 
means that the effect of anisotropy can be strong enough to counteract and even 
reverse the tendency of the field to enhance stabilization by extending the 
cut--off $k_s$ wave number beyond the conventional range set by Rayleigh.
\item
As the existence of magnetic anistropy is not limited to complex fluids such 
as LC's, the result of this work can be considered of a more general nature. 
\end{itemize}

\newpage
\appendix
\renewcommand{\theequation}{\thesection\arabic{equation}}
\section{Contribution of the magnetic field to free energy}
\label{appendix1}
\setcounter{equation}{0}
Evaluate the contribution $\mu_0 U/2$ of the magnetic field inside   
$\Omega_{cyl}$ and outside ${\mathbb R}^3\setminus \Omega_{cyl}$ of the 
disturbed liquid cylinder to the excess free energy $W$ 
\begin{eqnarray}
U&=&\chi_{\parallel}\left[
\int_{\Omega_{cyl}}\left(\;^{\sf in}H_z\right)^2dv-
\int_{\Omega_{cyl}^0}H^2dv \right]+\int_{\Omega_{cyl}}
\left[\left(\;^{\sf in}H_z\right)^2+
(1+\chi_{\perp})\left(\;^{\sf in}H_r\right)^2\right]dv+\label{contrib1}\\
&&2\chi_a \int_{\Omega_{cyl}}n_rn_z\;{}^{\sf in}H_r\;{}^{\sf in}H_zdv+
\int_{{\mathbb R}^3\setminus \Omega_{cyl}}
\left[\left(\;^{\sf ex}H_z\right)^2+\left(\;^{\sf ex}H_r\right)^2\right]dv-
\int_{{\mathbb R}^3}H^2dv\nonumber\\
&=&\chi_{\parallel}\left[
\int_{\Omega_{cyl}}\;\left(H_0+^{\sf in}\!\!H_z^1\right)^2dv-
\int_{\Omega_{cyl}^0}H_0^2dv\right]+\int_{\Omega_{cyl}}
\left[\;\left(H_0+^{\sf in}\!\!H_z^1\right)^2+
(1+\chi_{\perp})\left(\;^{\sf in}H_r^1\right)^2\right]dv+
\nonumber\\
&&2\chi_a \int_{\Omega_{cyl}}n_r n_z \left(H_0+\;^{\sf in}H_z^1\right)
\;^{\sf in}H_r^1 dv+\int_{{\mathbb R}^3\setminus \Omega_{cyl}}\;
\left[\left(H_0+^{\sf ex}\!\!H_z^1\right)^2+
\left(^{\sf ex}H_r^1\right)^2\right]dv-\int_{{\mathbb R}^3}H_0^2dv\nonumber\\
&=&\chi_{\parallel}H_0^2
\left(\int_{\Omega_{cyl}}dv-\int_{\Omega_{cyl}^0}dv\right)+
\int_{\Omega_{cyl}}\left\{(1+\chi_{\parallel})\left[2H_0\;^{\sf in}H_z^1+
\left(^{\sf in}H_z^1\right)^2\right]+
(1+\chi_{\perp})\left(^{\sf in}H_r^1\right)^2\right\}dv+\nonumber\\
&&2\chi_a H_0\int_{\Omega_{cyl}}n_r n_z \;^{\sf in}H_r^1 dv+
\int_{{\mathbb R}^3\setminus \Omega_{cyl}}
\left\{2H_0\;^{\sf ex}H_z^1+
\left(^{\sf ex}H_z^1\right)^2+\left(^{\sf ex}H_r^1\right)^2\right\}dv
\nonumber\\
&=&(1+\chi_{\parallel})\int_{\Omega_{cyl}}\left[2H_0\;^{\sf in}H_z^1+
\left(^{\sf in}H_z^1\right)^2\right]dv+
(1+\chi_{\perp})\int_{\Omega_{cyl}}\left(^{\sf in}H_r^1\right)^2dv+
\nonumber\\
&&2\chi_a H_0\int_{\Omega_{cyl}}n_r n_z \;^{\sf in}H_r^1 dv+
\int_{{\mathbb R}^3\setminus \Omega_{cyl}}
\left\{2H_0\;^{\sf ex}H_z^1+
\left(^{\sf ex}H_z^1\right)^2+\left(^{\sf ex}H_r^1\right)^2\right\}dv
\nonumber\\
&=&\int_{\Omega_{cyl}}\left\{(1+\chi_{\parallel})\left(^{\sf in}H_z^1\right)^2+
(1+\chi_{\perp})\left(^{\sf in}H_r^1\right)^2\right\}dv+
\int_{{\mathbb R}^3\setminus \Omega_{cyl}}
\left\{\;\left(^{\sf ex}H_z^1\right)^2+
\left(^{\sf ex}H_r^1\right)^2\right\}dv+
\nonumber\\
&&2H_0\left(\chi_a \int_{\Omega_{cyl}}n_r n_z \;^{\sf in}H_r^1 dv+
(1+\chi_{\parallel})\int_{\Omega_{cyl}}\;^{\sf in}H_z^1dv+
\int_{{\mathbb R}^3\setminus \Omega_{cyl}}\;^{\sf ex}H_z^1dv\right)\;.\nonumber
\end{eqnarray}
For the aims, discussed in section \ref{expen}, we also give the linear in 
$\chi_a$ representations
\begin{eqnarray}
U=U_0+\chi_a U_1\;,\;\;\;^{\sf in}H_r^1=\overline{^{\sf in}H_r^1}+
\widetilde{^{\sf in}H_r^1}\;,\;\;\;^{\sf in}H_z^1=\overline{^{\sf 
in}H_z^1}+\widetilde{^{\sf in}H_z^1},\label{contrib2}
\end{eqnarray}
where
\begin{eqnarray}
U_0&=&(1+\chi_{\perp})\int_{\Omega_{cyl}}\left[\left(\overline{^{\sf in}H_z^1}
\right)^2+\left(\overline{^{\sf in}H_r^1}\right)^2\right]dv+\int_{{\mathbb R}^
3\setminus \Omega_{cyl}}\left[\;\left(^{\sf ex}H_z^1\right)^2+\left(^{\sf ex}
H_r^1\right)^2\right]dv+\label{contrib2a}\\
&&2H_0\left((1+\chi_{\perp})\int_{\Omega_{cyl}}\;\overline{^{\sf in}H_z^1}dv+
\int_{{\mathbb R}^3\setminus \Omega_{cyl}}\;^{\sf ex}H_z^1dv\right)\;,
\nonumber\\
U_1&=&\frac{2(1+\chi_{\perp})}{\chi_a}\int_{\Omega_{cyl}}\left(
\overline{^{\sf in}H_z^1}\;\widetilde{^{\sf in}H_z^1}+\overline{^{\sf in}H_r^1}
\;\widetilde{^{\sf in}H_r^1}+H_0\;\widetilde{^{\sf in}H_z^1}\right)dv+
\label{contrib2b}\\
&&\int_{\Omega_{cyl}}\left[\left(\overline{^{\sf in}H_z^1}\right)^2+2n_rn_zH_0\;
\overline{^{\sf in}H_r^1}+2H_0\;\overline{^{\sf in}H_z^1}\right]dv.\nonumber
\end{eqnarray}
In the last formulas we introduced the following notations
\begin{eqnarray}
\overline{^{\sf in}H_r^1}=-kA_1I_1(kr)\sin kz\;,\;\;\;
\widetilde{^{\sf in}H_r^1}=-\chi_a\frac{A_{\phi}}{k}N_2I_1(kr)\sin kz\;,
\label{contrib3}\\
\overline{^{\sf in}H_z^1}=-kA_1I_0(kr)\cos kz\;,\;\;\;
\widetilde{^{\sf in}H_z^1}=-\chi_a\frac{A_{\phi}}{k}N_1I_0(kr)\cos kz\;.
\nonumber
\end{eqnarray}
In fact, $U_0$ was calculated in \cite{felzim03}
\begin{eqnarray}
U_0=-\pi L\left(\chi_{\perp}\zeta_0kRH_0\right)^2I_0(kR)K_0(kR)\;.
\label{contrib4}
\end{eqnarray}
Calculate the integrals in (\ref{contrib2b}) taking in mind $\chi_{\perp}\ll 1$
\begin{eqnarray}
\frac{1}{\chi_a}\int_{\Omega_{cyl}}\overline{^{\sf in}H_z^1}\;
\widetilde{^{\sf in}H_z^1}dv&=&A_1A_{\phi}N_1\int_{\Omega_{cyl}}I_0^2(kr)
\cos^2kzdv=\frac{\pi L}{2}A_1A_{\phi}N_1R^2\left[I_0^2(kR)-I_1^2(kR)\right]
\nonumber\\
&=&-\frac{\pi L}{2}\chi_{\perp}(\zeta_0H_0)^2(kR)^3N_1\left[I_0^2(kR)-
I_1^2(kR)\right]\frac{K_0(kR)}{I_1(kR)}\;,\label{integ1a}\\
\frac{1}{\chi_a}\int_{\Omega_{cyl}}\overline{^{\sf in}H_r^1}\;\widetilde{^{\sf 
in}H_r^1}dv&=&A_1A_{\phi}N_2\int_{\Omega_{cyl}}I_1^2(kr)\sin^2kzdv=
\frac{\pi L}{2}A_1A_{\phi}N_2R^2\left[I_1^2(kR)-I_0(kR)I_2(kR)\right]\nonumber\\
&=&-\frac{\pi L}{2}\chi_{\perp}(\zeta_0H_0)^2(kR)^3N_2\left[I_1^2(kR)-
I_0(kR)I_2(kR)\right]\frac{K_0(kR)}{I_1(kR)}\;,\label{integ1b}\\
\frac{H_0}{\chi_a}\int_{\Omega_{cyl}}\widetilde{^{\sf in}H_z^1}dv&=&
-\frac{A_{\phi}}{k}N_1H_0\int_{\Omega_{cyl}}I_0(kr)\cos kzdv=
\pi L\zeta_0R\frac{A_{\phi}}{k}N_1H_0I_0(kR)=\nonumber\\
&=&-\pi LN_1(\zeta_0H_0)^2kR\frac{I_0(kR)}{I_1(kR)}\;,\label{integ1c}\\
\int_{\Omega_{cyl}}\left(\overline{^{\sf in}H_z^1}\right)^2dv&=&
k^2A_1^2\int_{\Omega_{cyl}}I_0^2(kr)\cos^2kzdv=\frac{\pi L}{2}A_1^2(kR)^2
\left[I_0^2(kR)-I_1^2(kR)\right]\nonumber\\
&=&\frac{\pi L}{2}\left(\chi_{\perp}\zeta_0H_0\right)^2(kR)^4
\left[I_0^2(kR)-I_1^2(kR)\right]K_0^2(kR)\;,\label{integ1d}\\
H_0\int_{\Omega_{cyl}}n_r n_z\overline{^{\sf in}H_r^1}dv&=&-k^2A_1A_2H_0\int_
{\Omega_{cyl}}I_1^2(kr)\sin^2kzdv\nonumber\\
&=&-\frac{\pi L}{2}A_1A_2(kR)^2H_0\left[I_1^2(kR)-I_0(kR)I_2(kR)\right]\nonumber\\
&=&\frac{\pi L}{2}\chi_{\perp}(\zeta_0H_0)^2(kR)^3\left[I_1^2(kR)-I_0(kR)
I_2(kR)\right]\frac{K_0(kR)}{I_1(kR)}\;,\label{integ1e}\\
H_0\int_{\Omega_{cyl}}\overline{^{\sf in}H_z^1}dv&=&-kA_1H_0\int_{\Omega_{cyl}}
I_0(kr)\cos kzdv=\pi L A_1H_0\zeta_0kRI_0(kR)\nonumber\\
&=&\pi L \chi_{\perp}(\zeta_0H_0)^2(kR)^2I_0(kR)K_0(kR)\;.\label{integ1f}
\end{eqnarray}
It is quite surprising that among all above integrals there is only one 
(\ref{integ1c}) which dominates over the others in the region $\chi_{\perp}\ll 
1$. Thus, in this limit we finally have
\begin{eqnarray}
-U=\pi L(\zeta_0H_0)^2\chi_{\perp}kR\frac{I_0(kR)}{I_1(kR)}\left(\chi_{\perp}
kRI_1(kR)K_0(kR)+2N_1\frac{\chi_a}{\chi_{\perp}}\right)\;.\label{integ2}
\end{eqnarray}
The last expression shows that the anisotropic part of $U$ can prevail 
over the isotropic one provided that, 
\begin{eqnarray}
|N_1\chi_a|>\chi_{\perp}^2\;.
\label{integ2a}
\end{eqnarray}

\newpage
\section{On the solution of Equation (\ref{exp5}).}
\label{appendix2}
\setcounter{equation}{0}
The non--homogeneous equation
\begin{eqnarray}
\left(\frac{d^2 }{d x^2}+\frac{1}{x}\frac{d}{d x}-1\right)\Psi=CI_0(x)\;,
\;\;\;\Psi(x)=CG(x)I_0(x)\;,\;\;\;\Psi(x_0)=0\;,
\label{exp6}
\end{eqnarray}
leads to the non--homogeneous equation for the amplitude function $G(x)$
\begin{eqnarray}  
\frac{d^2 G}{d x^2}+\left(2\frac{I_1(x)}{I_0(x)}+\frac{1}{x}\right)
\frac{dG}{d x}=1\;,\;\;\;G(x_0)=0\;,
\label{exp7}
\end{eqnarray}
or, after substitution $dG/dx=S(x)$, essentially simplifies the problem
\begin{eqnarray}
\frac{dS}{d x}+\left(2\frac{I_1(x)}{I_0(x)}+\frac{1}{x}\right)S=1\;.
\label{exp8}  
\end{eqnarray}
Its solution reads
\begin{eqnarray}
S(x)&=&\exp\left(-\int\left(2\frac{I_1(x)}{I_0(x)}+\frac{1}{x}\right)dx\right)
\int\exp\left(\int\left(2\frac{I_1(t)}{I_0(t)}+\frac{1}{t}\right)dt\right)dy=
\nonumber\\
&&\frac{1}{xI_0^2(x)}\int yI_0^2(y)dy=
\frac{x}{2}\left(1-\frac{I_1^2(x)}{I_0^2(x)}\right)\;,\nonumber
\end{eqnarray}
and finally   
\begin{eqnarray}  
G(x)=\frac{1}{4}\left(x^2-x_0^2\right)
-\frac{1}{2}\int_{x_0}^x \frac{I_1^2(y)}{I_0^2(y)}ydy\;.
\label{exp10}
\end{eqnarray}
\end{document}